\documentclass[twocolumn,secnumarabic,amssymb, nobibnotes, aps, prl, superscriptaddress]{revtex4-1}

\usepackage{float}
\usepackage{graphicx}
\usepackage{subfigure}
\usepackage{amsmath} 
\usepackage{epstopdf}
\usepackage{setspace}
\usepackage[font=small,labelfont=bf,belowskip=-15pt,aboveskip=0pt]{caption}
\usepackage[shortlabels]{enumitem}
\usepackage{mathrsfs}
\usepackage{subfig}
\usepackage{sidecap}
\usepackage[export]{adjustbox}
\usepackage{enumitem}
\usepackage{color,soul}
\usepackage[dvipsnames]{xcolor}
\usepackage[sort&compress]{natbib}
\usepackage[normalem]{ulem}
\usepackage{dcolumn,amssymb,subfig}
\usepackage{bm}
\usepackage{wrapfig}
\usepackage{caption}
\usepackage[margin=1in]{geometry}
\usepackage{xcolor,colortbl}
\usepackage{multirow}
\usepackage{xfrac}

\newcommand{\del}{\boldsymbol{\nabla}}
\newcommand{\fig}[1]{\textbf{Fig.~\ref{#1}}}
\newcommand{\movie}[1]{\textbf{movie~#1}}
\newcommand{\movies}[1]{\textbf{movies~#1}}
\newcommand{\eq}[1]{\textbf{Eq.~\ref{#1}}}
\renewcommand{\vec}[1]{\boldsymbol{#1}}
\newcommand{\tens}[1]{\boldsymbol{#1}}
\newcommand{\av}[1]{\left\langle #1 \right\rangle}

\begin{document}
\title{Twist-induced crossover from 2D to 3D turbulence in active nematics}%

\author{Tyler N. Shendruk\textsuperscript{\ddag}}
\email{tshendruk@rockefeller.edu}
\affiliation{The Rockefeller University, 1230 York Avenue, New York, New York, 10021}
\thanks{These authors contributed equally}
\author{Kristian Thijssen\textsuperscript{\ddag}}
\affiliation{Rudolf Peierls Centre for Theoretical Physics, University of Oxford, Oxford OX1 3NP, UK}
\author{Julia M. Yeomans}
\affiliation{Rudolf Peierls Centre for Theoretical Physics, University of Oxford, Oxford OX1 3NP, UK}
\author{Amin Doostmohammadi}
\email{amin.doostmohammadi@physics.ox.ac.uk}
\affiliation{Rudolf Peierls Centre for Theoretical Physics, University of Oxford, Oxford OX1 3NP, UK}

\begin{abstract}
While studies of active nematics in two dimensions have shed light on various aspects of the flow regimes and topology of active matter, three-dimensional properties of topological defects and chaotic flows remain unexplored. 
By confining a film of active nematics between two parallel plates, we use continuum simulations and analytical arguments to demonstrate that the crossover from quasi-2D to 3D chaotic flows is controlled by the morphology of the disclination lines. 
For small plate separations, the active nematic behaves as a quasi-2D material, with straight topological disclination lines spanning the height of the channel and exhibiting effectively 2D active turbulence. 
Upon increasing channel height, we find a crossover to 3D chaotic flows due to the contortion of disclinations above a critical activity. 
We further show that these contortions are engendered by twist perturbations producing a sharp change in the curvature of disclinations. 
\end{abstract}

\maketitle

Active matter includes a wide range of biological and synthetic materials that are driven out-of-equilibrium by continuous energy injection from their internal elements \cite{Sriram2010,Marchetti2013,Bechinger2016}. 
The constituent particles are typically elongated, as exemplified by filamentous motor protein/microtubule bundles~\cite{Dogic2012,Francesc2016}, or motile bacilliform fluids~\cite{Volfson2008,Grosmann2016}. 
In dense suspensions, the {\em nematic} nature of the interactions between these particles results in orientational order, which is continuously disturbed by active stresses, leading to topological discontinuities. 

In {\em two-dimensional} active nematics, the discontinuities are point-like {\em topological defects}. Defects are an unavoidable consequence of broken continuous symmetry~\cite{mermin1979}.
Nematic defects have been reported in microtubule bundle films~\cite{Dogic2012,Keber2014,Dogic2015,Francesc2016}, in thin films of actin filaments~\cite{Zhang2017}, in sporadically direction-reversing bacteria and progenitor neural stem cells~\cite{Grosmann2016,Sano2017}, and even in cell populations that consist of polar-but-elongated individuals, such as fibroblast cells~\cite{Duclos2017}, and layers of epithelial cells~\cite{Saw2017}. 
{\it In-vivo} experiments have uncovered biological functionality of nematic defects governing cell death and extrusion of epithelial cells from monolayers~\cite{Saw2017}, and mound formation in progenitor neural stem cells~\cite{Sano2017,Charras2017}. 

\begin{figure*}[ht!]
{\centering
\includegraphics[width=0.95\linewidth]{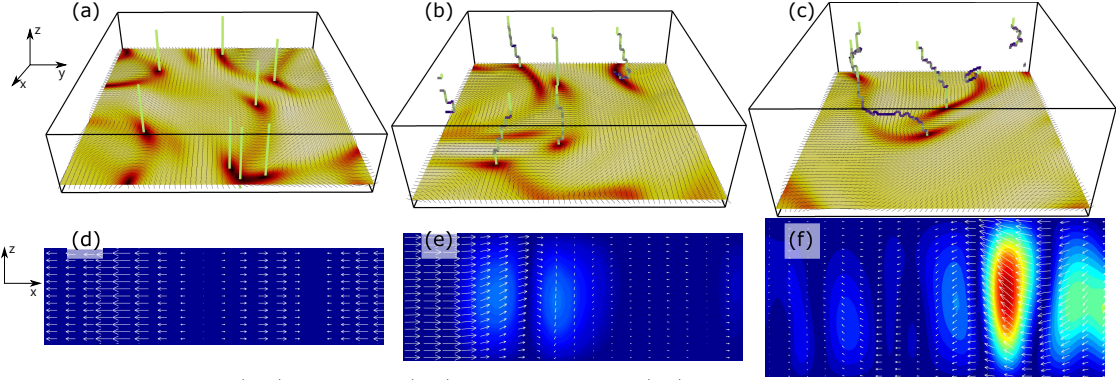}}
\caption{Snapshots of (a,d) quasi-2D, (b,e) transitional and (c,f) 3D confined active nematics. The dynamics change from a quasi-2D flow with straight disclination lines for the channel height $H=15$ ($A=15$) in (a) through a transition regime near $H=20$ ($A=20$) in (b) to 3D flows with strongly contorted disclination lines for the channel height $H=25$ ($A=25$) in (c). 
In the top row (a-c), the planar colourmap illustrates the magnitude of the nematic order $\mathcal{S}$ and director field $\vec{n}$ (solid black lines) in the vicinity of the lower bounding free-slip wall. 
Disclination lines are shown as thick lines coloured by the characteristic disclination angle $\alpha$, from wedge-type disclination segments with $\alpha=0$ (green) to twist-type segments with $\alpha=\pi/2$ (purple).  
The bottom row (d-f) shows a slice of the velocity fields for the same channel heights. 
The length of the arrow indicates the local speed, while the colourmap shows the cross-channel component $v_z$.
}
\label{fig:snaps}
\end{figure*}

The defect dynamics are intrinsically connected to the chaotic flows in 2D active fluids in the nematic phase. 
In contrast to passive nematic films, topological defect pairs are continuously created and annihilated~\cite{Giomi2013,ourepl2014,Giomi2015,Keber2014,Francesc2016}. 
The capacity of active nematics to maintain and organise the resulting steady state defect population has stimulated a recent surge of interest in defect dynamics in condensed matter systems~\cite{Zhang2017,Allahyarov2017,cortese2017}. 
In active nematics, the chaotic motion of the defects drives vortices and jets, generating the disorderly flow state of {\it active or mesoscale turbulence}~\cite{Kessler2004,Julia2012,Dogic2012,ourprl2013}. 
Previous theoretical, experimental, and numerical studies have explored the topology and flow characteristics of 2D mesoscale turbulence in active nematics~\cite{Giomi2015,Blanch17} and the flow features in 3D mesocale turbulence~\cite{Urzay2017,Slomka2017}. 
In an infinite 2D system, the ordered nematic state is hydrodynamically unstable to any active perturbation. 
However in a confined geometry, the chaotic flows of active nematics can be stabilised into spatiotemporally ordered flow states and defect trajectories~\cite{Francesc2016,Dogic2015,DoostmohammadiNC2016,ShendrukSM2017,Norton2018}. 

While extensive research has been dedicated to understanding active nematics and topological defects in such two-dimensional systems, basic defect properties and flow patterns are yet to be explored in three-dimensions. 
Indeed, recent experiments by Wu, {\it et al.}~\cite{Wu2017}  have demonstrated that three-dimensional confinement of isotropic active fluids can drive a transition from turbulent flow to a long-range coherent flow depending on the channel aspect ratio, showing that higher dimensionality can play a significant role in active fluid behaviour. 

As a step towards characterising active turbulence in three dimensions, we numerically study the crossover from 2D to 3D structures in an active system by considering an active nematic fluid confined between two parallel plates. 
When the spacing between the plates is small, we observe a `quasi-2D' regime where straight lines of topological disclinations span the system (\fig{fig:snaps}a; \movie{1}~\cite{SI}), which behaves as a stack of identical 2D layers. As the distance between the plates is increased there is a crossover to full 3D active turbulence, which we show is driven by the contortion of the disclinations (\fig{fig:snaps}b,c; \movies{2-3}~\cite{SI}).

To simulate the crossover from 2D to 3D active turbulence, we solve the nematohydrodynamic equations~\cite{Davide2007,Giomi2013} of motion for an active nematic confined between two parallel planar surfaces separated by a varying distance $H$. 
The nematic order is described by the tensor $\tens{Q}=3\mathcal{S} \left( \vec{n}\vec{n}-\tens{I}/3 \right) /2$, with director field $\vec{n}$ and scalar order parameter $\mathcal{S}$, which can vary from $\mathcal{S}_{eq}=1/3$ in ordered regions to zero at the core of topological defects. 
The nematic field evolves according to $D_t \tens{Q} - \tens{S} = \tens{H}/\gamma$~\cite{Berisbook}, which describes the relaxation of the orientation towards equilibrium at a rate determined by the rotational viscosity $\gamma=2.94$. 
The rate of change of $\tens{Q}$ is described by the material derivative $D_t$ and $\tens{S}$, the co-rotational advection of the nematic tensor due to gradients of the velocity field~\cite{Davide2007}, which includes the alignment parameter $\lambda=0.3$. 

The molecular field $\tens{H}$ includes the Landau-de~Gennes free energy, as well as distortion free energy density terms
\begin{align*}
f &= \frac{A}{2}Q_{ij}Q_{ji} + \frac{B}{3}Q_{ik}Q_{kl}Q_{lj} + \frac{C}{4}\left(Q_{ij}Q_{ji}\right)^2 +f_\text{el},\\
f_\text{el} &= \frac{L_1}{2}\partial_k Q_{ij} \partial_k Q_{ij} + \frac{L_2}{2}\partial_k Q_{kj} \partial_i Q_{ij}\\
 &\qquad + \frac{L_3}{2} Q_{ki} \partial_k Q_{jl} \partial_i Q_{jl} . 
\end{align*}
The Landau-de Gennes coefficients we use are $A=0$, $B=-0.3$ and $C=0.3$. 
The tensorial elastic constants $L_i$ are mapped to the Frank elastic constants $K_i$ through the relations~\cite{Trimper1983}
\begin{align*}
  \frac{K_\text{Bend}}{2\mathcal{S}_{eq}^2} &= L_1 + \frac{L_2}{2} - \frac{\mathcal{S}_{eq}L_3}{3} \\
  \frac{K_\text{Twist}}{2\mathcal{S}_{eq}^2} &= L_1 - \frac{\mathcal{S}_{eq}L_3}{3} \\
  \frac{K_\text{Splay}}{2\mathcal{S}_{eq}^2} &= L_1 + \frac{L_2}{2} - \frac{2\mathcal{S}_{eq}L_3}{3}. 
\end{align*}
Unless otherwise stated, we use the one-constant approximation for which only $L_1$ is non-zero and all values of $K_i=K$, which is varied in the range $\left[ 0.01,0.05 \right]$. 
When we later relax the one constant approximation, we make the simplifying choice $K_\text{Splay}=K_\text{Bend}=K$ in order to vary twist relative to the other elastic constants. 

The velocity field $\vec{u}$ obeys the incompressible Navier-Stokes equation $D_t \vec{u}=\del\cdot\tens{\Pi} / \rho$, in which the generalized stress $\tens{\Pi}$ has viscous ($\eta=2/3$), elastic, and active components~\cite{Davide2007B}. 
The active stress is described by $-\zeta\tens{Q}$~\cite{Sriram2002} such that the divergence of $\tens{Q}$ drives active forcing. 
Thus, not only does contortion of the disclination lines arise from stress, as is typical in traditional nematics, but also active forcing arises from contortions through the activity. 
This work focuses on extensile active fluids, relevant to microtubule/kinesin bundles~\cite{Dogic2012,Francesc2016}, for which the activity parameter $\zeta>0$ and in the range $\left[ 0.01,0.05 \right]$. 
The fluid density is taken to be constant with $\rho=1$. 
This active nematic model has been described in detail in previous publications.~\cite{ourprl2013,ourpta2014,ShendrukSM2017,DoostmohammadiNC2017}. 
Here, we do not need additional terms in the active stress that may be relevant when the confinement size is comparable to the size of the constituent active particles~\cite{maitra2017}. 
We choose simulation parameters in a range that reproduces flow patterns of 2D microtubule bundles under confinement~\cite{note2}. 
The impermeable parallel surfaces impose strong planar anchoring~\cite{Mottram2014} on the nematic field and free-slip boundary conditions on the velocity, except where otherwise stated. 

The active nematohydrodynamic equations are solved using a hybrid lattice Boltzmann and finite difference method~\cite{ourprl2013}. 
Simulations were performed in a cuboid of volume $100\times100\times H$. 
The planar channel geometry, characterized by the plate separation $H$, competes with the characteristic length scale of active turbulence $\sim\sqrt{K/\zeta}$ \cite{ourpta2014,Hemingway2016}, resulting in the {\em dimensionless activity number} $A = H\sqrt{\zeta/K}$. 

\begin{figure}
\centering
\includegraphics[width=1.0\columnwidth]{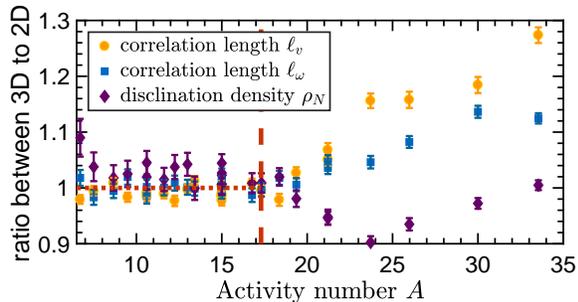}
\caption{Comparison of turbulence properties in 2D and 3D simulations. 
Shown here are (a) velocity correlation length $\ell_{v}$, (b) vorticity correlation length $\ell_{\omega}$, and (c) disclination number density $\rho_{N}$. 
All values are normalised by their 2D counterpart. 
The dashed red line marks the crossover from quasi-2D to 3D active turbulence.}
\label{fig:2D3D}
\end{figure}

When the channel height $H$ is sufficiently small compared to $\sqrt{K/\zeta}$, the active turbulence is quasi-2D. 
In this limit, both flow and director fields are height-independent, and topological defects form straight disclination lines normal to the surfaces that directly span the gap with translational invariance across the channel (\fig{fig:snaps}a; \movie{1}~\cite{SI}). 
When observed from above, the disclination lines appear as 2D point defects with half-integer charges $m = \pm 1/2$. 
Disclinations are continuously created and annihilated, such that in every plane parallel to the channel walls the defect dynamics is effectively that of 2D active turbulence~\cite{Giomi2013,ourepl2014}. 

The flow fields reflect this quasi-2D behaviour for sufficiently small channel heights, showing 2D active turbulence in $xy$-planes parallel to the wall but translationally invariant across the channel (\fig{fig:snaps}d; \movie{4}~\cite{SI}). 
The turbulent flow dynamics are quantified through the velocity and vorticity correlation lengths, as well as the number density of disclinations (\fig{fig:2D3D}). 
The velocity and vorticity length scales are calculated from the the velocity-velocity correlation function $C_{vv}(r)=\langle\vec{u}(r,t)\cdot\vec{u}(0,t)\rangle / \langle\vec{u}(0,t)^2\rangle$ and vorticity-vorticity correlation function $C_{\omega\omega}(r)=\langle\vec{\omega}(r,t)\cdot\vec{\omega}(0,t)\rangle / \langle\vec{\omega}(0,t)^2\rangle$, respectively. 
Our measurements in the observed quasi-2D turbulence reveal that the correlation lengths and number density of defects are all consistent with 2D behaviour up to a given channel heights. 
However, at a critical activity number, all three characterizations of the flow deviate from their 2D values. 

Although the active turbulence can be described as effectively two dimensional in this limit, indications of the 3D nature of the film remain apparent in the pair-production process. 
In quasi-2D active turbulence pair-production of effective 2D $\pm 1/2$ defects occurs when a 3D disclination loop forms in the centre of the channel (\movie{7}~\cite{SI}). 
The small ring inflates until it makes contact with the bounding planar surfaces, at which point it then splits into two straight disclination lines that bridge the gap and appear as the pair of $\pm 1/2$ defects in the xy-plane (\fig{fig:snaps}a). 
Apparent pair-production is a rapid process ($\sim10$ simulation time steps) and pair-annihilation events of oppositely charged apparent defects occur analogously, or as horseshoe shaped arches when the annihilating disclination lines make contact at one surface but not the other (\movie{7}~\cite{SI}). 

\begin{figure}
\includegraphics[width=1.0\columnwidth]{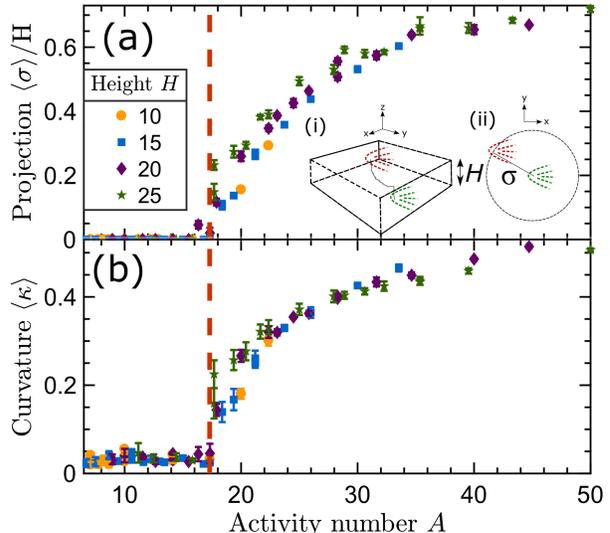} 
\caption{(a) The average projected distance $\langle\sigma\rangle$ between the ends of the disclinations for free-slip walls as a function of activity number $A$. 
\textit{Insets:} Schematic representation of $\sigma$. (b) Mean curvature $\av{\kappa}$ as a function of $A$. 
}
\label{fig:sigma}
\end{figure}

As the channel height is increased, the system crosses over from quasi-2D to fully 3D active turbulence. 
At the onset of the 3D behaviour, the disclination lines begin to contort (\fig{fig:snaps}b; \movie{2}~\cite{SI}) and the translational invariance of the velocity profile across the channel is lost (\fig{fig:snaps}e; \movie{5}~\cite{SI}). 
Contortion typically occurs near the centre of the channel, while the disclination lines are essentially 2D defects normal to the boundary near the walls (\fig{fig:snaps}b-c). 
The projected positions of the disclination points at the top and bottom walls are separated by a non-zero in-plane distance $\sigma$ (\fig{fig:sigma}a-{\it inset}). 
Plotting the ensemble-average of this distance $\av{\sigma}$ against the dimensionless activity number $A$ shows a sharp crossover at a critical activity number 
(\fig{fig:sigma}a), after which the average separation rises continuously from zero as the disclination lines contort. 
Because this active system is intrinsically out-of-equilibrium, we must be mindful not to think of this transformation of the flow dynamics in terms of a phase transition under equilibrium conditions but rather a dynamic crossover. 

\begin{figure}
\includegraphics[width=0.85\linewidth]{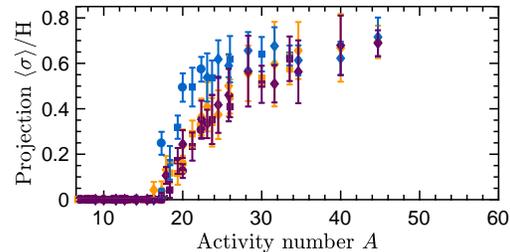}
\caption{The average projected distance $\langle\sigma\rangle$ between the ends of the disclinations as a function of activity number $A$ for various boundary conditions. 
 Walls with free-slip/strong-anchoring (yellow symbols) and free-slip/free-anchoring (purple) are indistinguishable, but no-slip/strong-anchoring (blue) rises more rapidly. 
 Three channel heights are used ($H=10$: circles; $15$: squares; $20$: diamonds).} 
\label{fig:slipAnchoring}
\end{figure}

For simulations with no-slip boundary conditions (\fig{fig:slipAnchoring}), the transition in the separation of the projected positions of the disclination lines is seen to be sharper but the critical activity is unchanged. 
This suggests that the nature of the instability that drives the transition from 2D to 3D turbulence is independent of the surface friction and that the breaking of translational symmetry across the channel gap by the no-slip condition causes the transition to rise more abruptly.
Likewise the nature of the crossover appears to be unaltered when the simulations do not impose any specific anchoring for the director at the boundaries but rather have free anchoring (\fig{fig:slipAnchoring}). 
It was previously shown that the extensile activity generates an effective planar active anchoring~\cite{MatthewPRL}. 

In addition to the projected distance between the disclination ends, measurements of the ensemble-average of the mean curvature of the disclination lines also show a sharp transition at the critical activity number (\fig{fig:sigma}b). 
This suggests that at the critical activity number, energy injected into the system around disclinations can overcome the elastic energy of the disclination line and therefore a finite curvature develops along the disclination. 
At this point, the cross-channel translational symmetry of the flow ceases and 3D active turbulence emerges.
Interestingly, both the normalised projected distance and curvature measurements for various channel heights collapse when plotted against the activity number, indicating that a constant critical dimensionless activity number $A_{\text{cr}}=[ H\sqrt{\zeta/K} ]_{\text{cr}}$ characterises the threshold between already established quasi-2D turbulence, with bend/splay deformations and in-plane flows, and fully 3D active dynamics. 

In what specific manner does this disclination distortion occur? 
In 2D films, the deformation of the director field around point defects is set by the splay and bend elastic constants (here assumed to be equal). 
However in 3D, twist distortions become possible. 
~Therefore to investigate the role of twist in the micro-structure of the disclination lines, we locally classify the disclination-type along each singularity's length. 
The characteristic disclination angle $\alpha$ is calculated~\cite{Hobdell97}. 
This differentiates between pure wedge-type disclinations ($\alpha=0$), which involve only splay and bend distortions, and pure twist-type disclinations ($\alpha=\pi/2$)~\cite{chandrasekhar1980,oswald2005}. 
The variation of $\alpha$ along the disclinations is shown in \fig{fig:snaps} and \movies{1-3}~\cite{SI}. 

\begin{figure}[tb]
{\centering
\includegraphics[width=1.0\columnwidth]{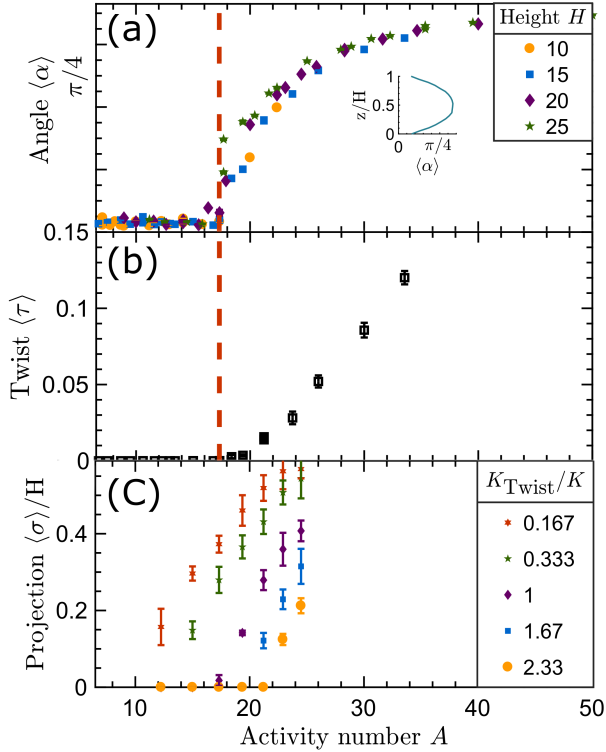}}
\caption{Contortion of disclinations are highly correlated to twist deformations. 
(a) Characteristic defect angle $\alpha$ as a function of activity number $A$. 
 \textit{Insets:} Profile along the defects across the channel for $A=20$. 
(b) Global mean twist deformation.  
(c) Increasing the twist elastic constant suppresses the contortion of defect lines. 
 Here, the activity number $A$ is defined using the bend/splay Frank coefficient $K$. 
}
\label{fig:twist}
\end{figure}

Below the critical activity number, the angle averaged over the contour length of all disclination lines $\av{\alpha}$ is close to zero (\fig{fig:twist}a), verifying that in the quasi-2D limit the disclination lines are wedge-type with bend/splay distortions confined to $xy$ planes. 
Beyond the critical activity number, however, the curves increasingly transform into twist-type disclinations. 
The average profile of the disclination angle $\alpha$ at each height through the channel shows a signature of nearly pure wedge-type at the walls and commonly twist-type in the mid-region  (\fig{fig:twist}(a)-\textit{inset}; \movie{2}~\cite{SI}). 
As the activity number is increased, the segment of twist-type disclination near the centreline broadens. 

To provide additional evidence of the role of twist, we calculate the system-wide mean twist deformation~\cite{Zumer2013}
\begin{equation*}
\av{\tau} = \frac{1}{\mathcal{S}^2}\left(\epsilon_{ikl} Q_{ij}\frac{\partial Q_{lj}}{\partial x_{k}} \right)^2.
\end{equation*} 
The mean twist becomes non-zero at the transition (\fig{fig:twist}b), where we find that twist is predominantly localised around disclination lines. 
However, deep in the fully 3D active turbulence regime, the twist deformation increases with activity number throughout the whole domain. 

To further check the role of twist in the contortion of disclinations and the subsequent crossover to 3D active turbulence, we suppress the transition by increasing the resistance of the active nematics to twist deformations. 
To this end, we relax the one-constant approximation for the orientational elasticity and progressively increase the twist coefficient $K_{\text{Twist}}$, while keeping the bend/splay constants equal to $K$. 
By increasing the twist elastic constant, the contortion of the disclinations is hampered; consequently, the crossover to 3D is retarded (\fig{fig:twist}c). 

These results suggest that the onset of 3D active nematic turbulence arises from the competition of energies. 
It is expected that a disclination line maintains its straight configuration when the elastic energy dominates, but contorts when the active energy injection overcomes the elastic energy barrier. 

The director configuration around a disclination line in the quasi-2D limit is $\vec{n}= \left[ {\cos}\left(\theta/2\right),{\sin}\left(\theta/2\right),0 \right]$~\cite{DeGennesBook}, where the factor $+1/2$ is the topological charge of the disclination, and $\theta$ is the polar angle in the $xy$-plane in the reference frame of the disclination. 
Adding a small twist perturbation $q$ per unit length along the initially straight line increases both the active energy that is dissipated in the system $E_\text{Act}$ and the elastic deformation energy $E_\text{Twist}$, which resists additional twist deformations. 
At the onset of the crossover from 2D to 3D behaviour, the director field at height $z$ can be represented as 
\begin{align}
 \vec{n} &= \left[ \cos\left(\frac{\theta+qz}{2}\right), \sin\left(\frac{\theta+qz}{2}\right) , 0 \right].
 \label{eq:n}
\end{align}
This twist creates a restoring elastic energy 
\begin{align}
 E_\text{Twist} &= \int \frac{K_{\text{Twist}}}{2}\left[\vec{n}\cdot \nabla\times \vec{n}\right]^2dV = \frac{\pi K_{\text{Twist}} R^2 q^2 H}{8}, 
\end{align}
where we have integrated over the cylindrical volume $V$ that is deformed by the presence of the disclination. 
This extends radially from the defect core to a range $R$, which is set by the average distance between disclinations~\cite{Giomi2015}. 

\begin{figure}[t]
\centering
\includegraphics[width=1.0\linewidth]{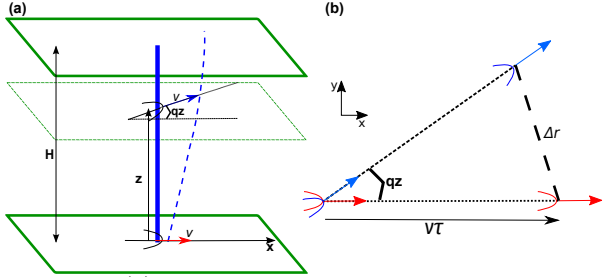}
\caption{(a) Schematic of a twisted disclination line. The solid line is the initial straight line and the dashed line represents the contorted disclination after time $\tau$. (b) Top view of the bottom disclination segment at height $z=0$ (red) and the top disclination segment at height $z$ (blue). The different segments of the disclination line move with a self-propulsion speed $v$ over a time interval $\tau$ (along the dashed line) at an angle $qz$ to the $x$-axis.
}
\label{fig:schematic}
\end{figure}

Due to the twist perturbation, different segments of the disclination move in different directions as a result of their self-propulsion (\fig{fig:schematic}). 
The relative displacement of a segment at height $z$ with respect to the bottom of the disclination is thus set by the amount of twist at height $z$ and, for  small twist $q\ll 1$, can be expressed as
\begin{equation}
  \Delta\vec{r}=
    \begin{bmatrix}
      v\tau\left(1-\cos{(qz)}\right)\\
      v\tau \sin{(qz)}\\
      0
    \end{bmatrix}
  \approx
    \begin{bmatrix}
      v\tau q^2z^2/2\\
      v\tau qz\\
      0
    \end{bmatrix}. 
    \label{eq:distortion}
\end{equation}
The magnitude of the displacement $v\tau$, is proportional to the disclination's self-propulsion speed $v\sim\zeta R/\eta$~\cite{Giomi2014} and the elastic relaxation time of the director field $\tau\sim H^2 \gamma/K$~\cite{Hemingway2016}. 
Therefore, $v\tau\sim RA^2\gamma/\eta$, where $\gamma/\eta$ is the ratio of the rotational viscosity of the nematic to the dynamic viscosity. 
In these simulations $\gamma/\eta\approx9/2$. 
The distortion (\eq{eq:distortion}) causes the polar angle in the $xy$-plane to be $\theta=\tan^{-1}\left(\left(y-\Delta r_{y}\right)/\left(x-\Delta r_{x}\right)\right)$. 

By integrating the active force density $-\zeta  \nabla \cdot \tens{Q}$ over a cylinder of radius $R$ and height $H$ around a disclination with orientation $\theta$, using \eq{eq:n}, the active force per unit length is found to be $\vec{F_{\text{Act}}} \approx \pi\zeta R\left[\left(-1+q^{2}z^{2}/2\right)~\hat{\vec{x}} + qz~\hat{\vec{y}}\right] /2 $, where $\hat{\vec{x}}$ is taken to lie in the direction of motion of the bottom point of the disclination line at $z=0$ and $\hat{\vec{y}}$ is orthogonally in-plane (see \fig{fig:schematic}).
We assume that the interaction range is much greater than the defect core and so  neglect any contribution from the defect core. 
When the twist perturbation is zero, we recover the known active force around a 2D point defect that results in self-motility of the $+1/2$ defects~\cite{Giomi2014}. 
Thus, the first term in the $\hat{\vec{x}}$-component of $\vec{F}_\text{Act}$ corresponds to the normal 2D deformation of the director field around a $+1/2$ disclination and the additional terms are due to the twist-induced deformations. 
Defining the active energy as the active force times the displacement and integrating over the length of disclination, we find the active energy dissipated within the fluid to be
\begin{align}
 E_\text{Act} &= \int_V \vec{F_{Act}}\cdot \vec{\Delta r}  dV \sim  \frac{\pi}{6} \zeta R v\tau q^{2}H^3. 
\end{align}

\begin{figure}[h]
{\centering
\includegraphics[width=1.0\columnwidth]{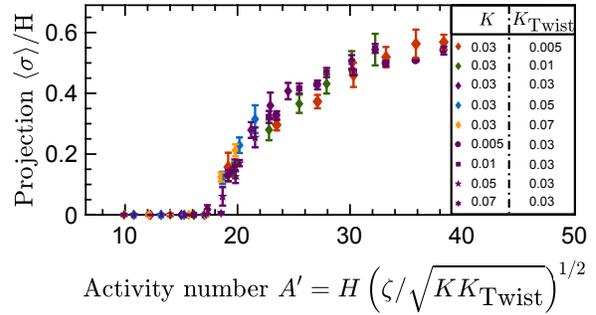}}
\caption{The average projected distance $\langle\sigma\rangle$ between the ends of the disclinations as a function of modified activity number $A^\prime$ for varying twist $K_{\text{Twist}}$ and bend/splay $K$ elastic constants. 
}
\label{fig:twistCollapse}
\end{figure}

When the active energy overcomes the restoring elastic deformation energy, twist perturbations grow and the system will not return to a quasi-2D  state.
Considering these energies, leads to a modified activity number 
\begin{align}
A^\prime &= H\left(\frac{\zeta}{\sqrt{KK_\text{Twist}}}\right)^{1/2} 
\end{align}
that reduces to $A=H\sqrt{\zeta/K}$ in a one-constant approximation $K_\text{Twist}=K$. 
At the crossover, the two competing energies must be comparable and we find 
\begin{align}
A^\prime_\text{cr} &\sim \left(\frac{\eta}{\gamma}\right)^{1/4}. 
\end{align}
Indeed measuring the projection distance $\sigma$ as a function of this modified activity number for varying the twist and bend/splay elastic constants, we find that the data collapse on a single curve (\fig{fig:twistCollapse}). 
This shows, in agreement with the theory, that the dimensionless activity number is the control parameter for the transition. 
Furthermore, it illustrates how the crossover to 3D active turbulence is induced by the competition between distorting active energy injection and the restoring twist elastic energy of the disclination.

Our results uncover a new physical mechanism in active nematics, showing that the crossover from quasi-2D to confined 3D active nematic turbulence is governed by the contortion of disclination lines above an activity threshold. 
The role of twist is specific to active 3D systems and thus our results suggest that future 3D active materials will exhibit rich physical dynamics not previously seen in either passive 3D nematics or 2D active monolayers.
\section*{Acknowledgement}
We would like to thank Paul van der Schoot for helpful discussions. KT was funded by the European Union's Horizon 2020 research and innovation programme under the Marie Sk\l{}odowska-Curie grant agreement No 722497. AD was supported by a Royal Commission for the Exhibition of 1851 Research Fellowship. 

\pagestyle{plain}

\end{document}